\documentclass[copyright,creativecommons]{eptcs}
\usepackage{amsmath,amssymb,amsthm}
\usepackage{graphicx}
\usepackage{alltt}
\usepackage{subfig}
\usepackage{boxedminipage}
\usepackage{afterpage}

  \newcommand{\arrow}{\longrightarrow}
\renewcommand{\partial}{\rightharpoonup}

\newcommand{\comment}[1]{}

\newcommand{\eg}{\textit{e.g.,}}
\newcommand{\ie}{\textit{i.e.,}}

\newcommand{\etc}{\textit{etc.}}

\newcommand{\E}{\mathcal{E}}
\newcommand{\N}{\mathbb{N}}

\graphicspath{./figs}
\providecommand{\event}{RTRTS 2010} 

\DeclareMathOperator{\rd}{rd}
\DeclareMathOperator{\src}{src}

\title{Using the PALS Architecture to Verify a Distributed Topology Control
Protocol for Wireless Multi-Hop Networks in the Presence of Node Failures
\thanks{This research was supported in part by the Boeing Corporation, grant C8088-557395, and by the NSF, grant CNS 08-34709}
}
\author{
  Michael Katelman and Jos\'{e} Meseguer
  \institute{Department of Computer Science \\
             University of Illinois at Urbana-Champaign \\
             Urbana, IL 61801, USA}
  \email{\{katelman,meseguer\}@uiuc.edu}
}

\def\titlerunning{Using PALS}
\def\authorrunning{Michael Katelman and Jos\'{e} Meseguer}

\begin{document}
\maketitle

\begin{abstract}
The PALS architecture reduces distributed, real-time asynchronous system design
to the design of a synchronous system under reasonable requirements. Assuming
logical synchrony leads to fewer system behaviors and provides a conceptually
simpler paradigm for engineering purposes. One of the current limitations of
the framework is that from a set of independent ``synchronous machines'', one
must compose the entire synchronous system by hand, which is tedious and
error-prone. We use Maude's meta-level to automatically generate a synchronous
composition from user-provided component machines and a description of how the
machines communicate with each other. We then use the new capabilities to
verify the correctness of a distributed topology control protocol for wireless
networks in the presence of nodes that may fail.
\end{abstract}

\section{Introduction}

The design and verification of a distributed embedded system (DES) such as
those in avionics, cars, medical systems, and sensor networks, poses serious
challenges for formal verification, particularly for model checking
verification,  for at least two reasons: (i) their real-time nature has to be
taken into consideration in their modeling and verification, and this usually
makes verification harder or may require restrictions such as the use of
time-bounded properties; (ii) their distributed nature can easily cause an
explosion in the size of the state space, making it infeasible for a model
checker to verify a system.

The Physically Asynchronous but Logically Synchronous (PALS) architectural
pattern \cite{meseguer_2010_01,meseguer_2009_01,sha_2009_01,miller_2009_01} has been recently
introduced to greatly reduce the design, verification, and implementation
efforts for a large class of DES systems, including many in avionics
applications, which can be conceptually conceived to operate in a synchronous
way, but which are in fact implemented as asynchronous systems.  Up to now, the
design of such systems has been very labor-intensive and error-prone, and their
formal verification has been infeasible due to state space explosion even for
modest-sized systems.  The essence of the PALS idea is to allow the designer
and formal verifier to specify the system as a \emph{synchronous composition of
abstract machines}, and to then \emph{automatically derive} from this
synchronous design a corresponding asynchronous version which is correct by
construction.

Conceptually, PALS can be understood as a \emph{model transformation},  which
takes as arguments both the simpler synchronous design and a collection of
\emph{performance-related upper bounds}, such as the maximum clock skew in an
underlying clock synchronization algorithm, the maximum network delay for
message transmission between any two components, and the maximum computation
time for an abstract machine to perform a one-step transition.  The result of
the PALS transformation is the corresponding asynchronous system that will
correctly realize the synchronous design with a provable minimal time period of
operation.

For model checking verification purposes, the great advantage of using PALS is
that the two difficulties (i)--(ii) described above for DES model checking
verification are greatly reduced.  Difficulty (i), their real-time nature,
completely goes away, since the synchronous system can be viewed as a
\emph{single abstract machine} obtained by composing the different machines
connected together in the design.  Such a single abstract machine, together
with its environment specification, can then be treated as an ordinary Kripke
structure \emph{with no explicit real-time features}, and can therefore be reasoned
about with standard model checkers.  Difficulty (ii) is enormously reduced,
because what in the synchronous model corresponds to a \emph{single state
transition} is achieved in its asynchronous version by possibly many transitions
(at least one per distributed component), with possibly many interleavings.
This means that the synchronous model will typically have many fewer states
than the asynchronous one, so that it may be often possible to model check the
synchronous model while it is infeasible to do so for its asynchronous version.
Because of the \emph{semantic equivalence} between a synchronous design and its
PALS asynchronous transform, both systems satisfy the same temporal logic
properties, so that it is enough to verify the much simpler synchronous design.

For the moment, the potential of PALS has not yet been fully exploited at the
formal specification and verification levels in languages such as Maude and
Real-Time Maude.  For this to happen, two important \emph{theory
transformations} need to be supported and automated, namely: 
\begin{enumerate}
\item the transformation performing the \emph{synchronous composition} of a
collection of abstract machines as specified by a \emph{wiring diagram}
connecting those machines; and

\item the PALS transformation itself, which takes the collection of such
machines and their wiring together with the performance parameters and produces
the equivalent asynchronous model.  
\end{enumerate} 
Thanks to the reflective features of rewriting logic and their Maude support by
its \texttt{META-LEVEL} module, transformations (1) and (2) can be specified
within Maude as reflective module transformations.  Note, however, that for
model checking verification purposes, since only the synchronous model needs to
be verified, \emph{only transformation} (1) \emph{is needed}.

This paper addresses this need for supporting PALS in Maude and Real-Time Maude
and makes the following contributions:
\begin{itemize}
\item It provides a meta-level  implementation in Maude of transformation (1)
which is both parametric on the wiring diagram and generic on the actual
abstract machines that may then be composed according to the specified diagram.

\item It applies the PALS transformation for the first time to an application
in the area of sensor networks, illustrating how it can be used to greatly
simplify the formal verification of  a topology control protocol (LMST)
\emph{in the presence of failures}, so that some of the nodes may fail.

\item It illustrates by the LMST example a more general method by which
real-time distributed object-based systems can be modeled in a much simpler
synchronous way using the PALS architecture, provided that the objects in the
system in question are supposed to \emph{only communicate with each other at
pre-established times}, and change their state at those times only as a result
of the messages they then receive.  
\end{itemize}

The remainder of the paper is organized as follows. Section \ref{sec:pals}
reviews the PALS framework, in particular the synchronous model of design.
Section \ref{sec:sc} then describes our generic implementation in Maude for
combining many independent synchronous machines into one large machine,
accomplishing transformation (1) above. In Section \ref{sec:lmst} we then
describe the LMST topology control protocol, our modeling of it in PALS using
the methods of Section \ref{sec:sc}, and prove its correctness with respect to
node failure using Maude's LTL model checker. Finally, Section
\ref{sec:conclusions} contains some concluding remarks and discusses future
work.

\section{Background: The PALS Synchronous Model}
\label{sec:pals}

To design a distributed, real-time system with the PALS architecture
\cite{meseguer_2010_01,meseguer_2009_01,sha_2009_01,miller_2009_01}, one starts with a
\emph{synchronous} model of the system; then, a very general
\emph{transformation}, formally specified in \cite{meseguer_2010_01,meseguer_2009_01}, takes the
synchronous model into an asynchronous one suitable for deployment. A
certain kind of \emph{bisimulation} between the two systems (see
\cite{meseguer_2010_01,meseguer_2009_01}) allows one to reduce (a) verification of a property
against the asynchronous machine to (b) verification of the property against
the synchronous design; where the properties in question are given by temporal
logic formulae. The synchronous machine typically has a much smaller state
space and can therefore be model checked more efficiently.

\begin{figure}
\begin{center}
\includegraphics[scale=0.5]{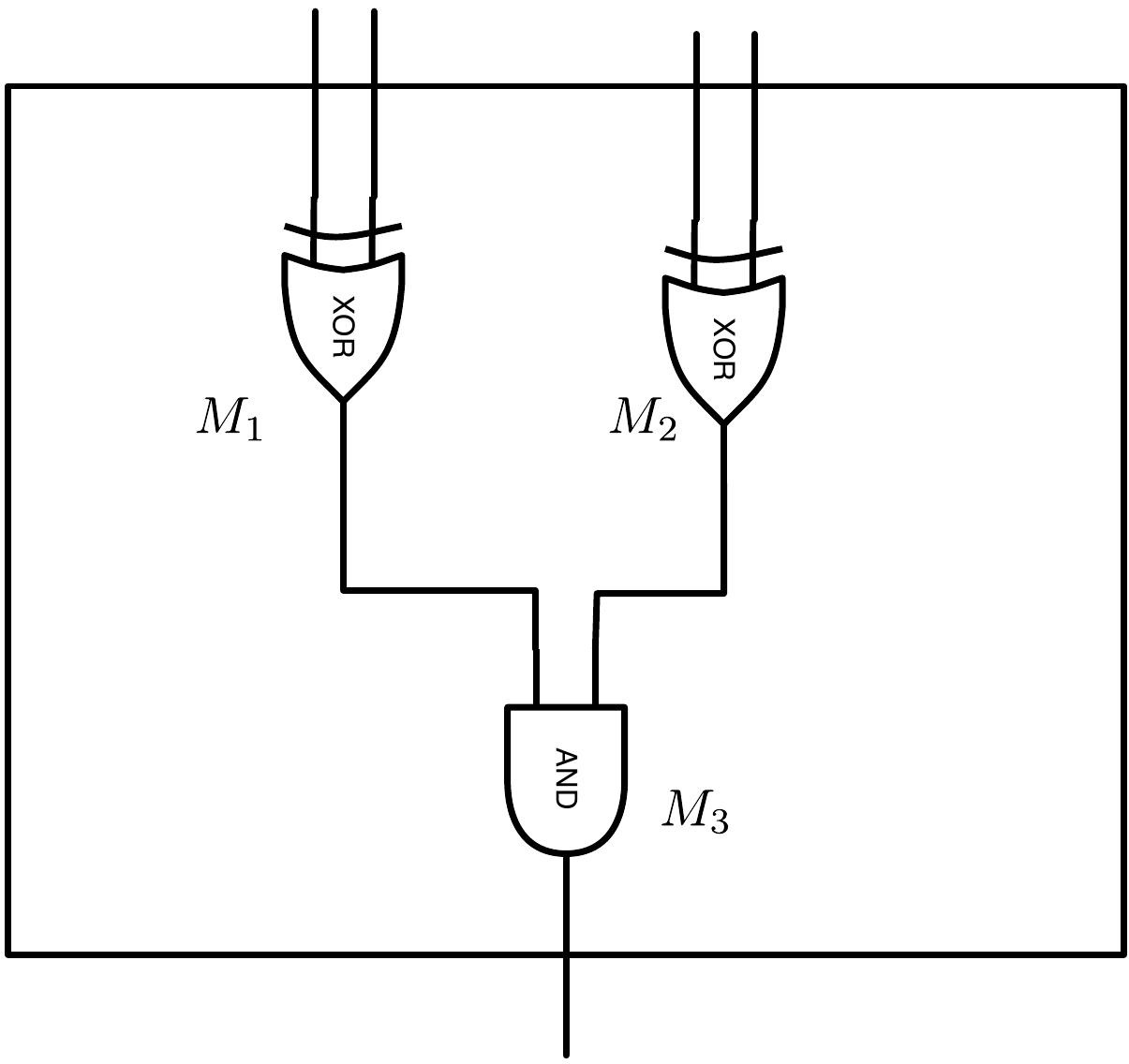}
\end{center}
\caption{A simple logic circuit.}
\label{fig:circuit}
\end{figure}

In this paper we are concerned only with the \emph{synchronous model} (for
information on the asynchronous model and transformation, see
\cite{meseguer_2010_01,meseguer_2009_01}), the key notions of which are \emph{synchronous
machines}, \emph{environment}, and \emph{wiring diagram}. Consider the small
logic circuit given in Figure \ref{fig:circuit}. In terms of the synchronous
model, it is comprised of three synchronous machines, $M_1$ -- $M_3$, an
environment defined by the unconnected wires at the boundary box, and a wiring
diagram specifying that the two xor gates get their inputs from the environment
and send their outputs to the and gate, which finally outputs its result to the
environment.

Formally, a component machine $M$ is defined as a four-tuple, $(D_i, S, D_o,
\delta : D_i \times S \arrow S \times D_o)$, where $D_i = D_{i_1} \times \dotsb
\times D_{i_n}, n \geq 1$, specifies the inputs to the machine, $D_o = D_{o_1}
\times \dotsb \times D_{o_m}, m \geq 1$, specifies the outputs of the machine,
$S$ is its internal state, and $\delta$ is its transition function, specifying
how the state is updated and what outputs are produced from a given input and
current state. Each of the component input types $D_{i_1},\dotsc,D_{i_n}$ and
component output types $D_{o_1},\dotsc,D_{o_m}$ are assumed to be non-empty; for
technical reasons, it is also assumed that $S \neq \emptyset$.

An environment is a pair $(D^e_i,D^e_o)$ with $D^e_i = D^e_{i_1} \times \dotsb
\times D^e_{i_{n_e}}, n_e \geq 1$, the set of inputs \emph{to} the environment,
and $D^e_o = D^e_{o_1} \times \dotsb \times D^e_{o_{m_e}}, m_e \geq 1$, the output
\emph{from} the environment.  Again, each of the
$D^e_{i_1},\dotsc,D^e_{i_{n_e}}$ and $D^e_{o_1},\dotsc,D^e_{o_{m_e}}$ are
assumed to be non-empty.

Let $\{M_j\}_{j \in J}$ be a ($J$) indexed set of machines and $E =
(D^e_i,D^e_o)$ an environment. A \emph{wiring diagram} for $\{M_j\}_{j \in J}$
and $E$ is a function $\src : I \arrow O$, where 
\begin{align*}
I &= \{ (j,n) \in (J \cup \{e\}) \times \mathbb{N} \mid 1 \leq n \leq n_{j}\} \qquad \text{and} \\
O &= \{ (j,n) \in (J \cup \{e\}) \times \mathbb{N} \mid 1 \leq n \leq m_{j}\},
\end{align*}
that maps each input port to the output port, or \emph{source}, from which it
receives data. Ports exist as part of the component machines, $\{M_j\}_{j \in
J}$, or as part of the environment, $E$.

Returning to the example in Figure \ref{fig:circuit}, each of the machines
$M_1$ -- $M_3$ have as input set $\mathbb{B}^2$, and as output set $\mathbb{B}$;
they have no internal state, but because the $S$ component is required to be
non-empty we represent the internal state with the unit type, we we denote by $\{\star\}$.
The output part of their transition functions perform exclusive-or in the case
of $M_1$ and $M_2$, and logical and in the case of $M_3$. The input type of the
environment is $\mathbb{B}$, since there is one input to the environment, and
output type $\mathbb{B}^4$, since the environment furnishes four values to the
machine. The wiring diagram, $\src$, is given by
\begin{align*}
(1,1) \mapsto (e,1), && (1,2) \mapsto (e,2), && (2,1) \mapsto (e,3), \\ 
(2,2) \mapsto (e,4), && (3,1) \mapsto (1,1), && (3,2) \mapsto (2,1), \\
(e,1) \mapsto (3,1).
\end{align*}
The composed machine operates just like the logic circuit, with values
propagating through one level of logic gate per round. This could, for example,
represent the time it takes for values to propagate through the logic
elements to the output of the circuit.

\section{Automatic Synchronous Composition}
\label{sec:sc}

We now describe the infrastructure we have built in Maude to compose a set of
synchronous machines into one larger machine, as illustrated above in Section
\ref{sec:pals} when we composed the three logic gates into a circuit. Given (a)
for each machine $M_j$, values $n_j,m_j$ specifying the number of inputs and
outputs, respectively, of $M_j$, (b) values $n_e,m_e$ for the number of inputs
and outputs of the environment, and (c) a wiring diagram, we automatically
generate a \emph{parameterized module} \cite[Ch. 10]{clavel_2007_01}
corresponding to the synchronous composition of a set of abstract ($\delta$
unspecified) machines composed according to the given wiring diagram. This is
accomplished using Maude's \emph{meta-level} \cite[Ch. 14]{clavel_2007_01}. The
module has a fixed topology, but is \emph{generic} in the operation of the
individual synchronous machines. The discussion in this section assumes a firm
understanding of parameterized and meta-level programming in Maude (see
\cite{clavel_2007_01} for a review).

Parameterized programming in Maude uses the notions of ``\emph{theories}'' and
``\emph{views}'' (see \cite[\S 8.3.1]{clavel_2007_01} and \cite[\S 8.3.2]{clavel_2007_01}). Theories are
used to specify a parameter's interface, and views are used to instantiate an
interface. Theories are like regular functional and system modules in Maude,
except that they do not need to satisfy the same conditions for executability.
In general, they also may omit constructors for defined sorts or equations
defining the declared operators since these will be mapped to other sorts and
functions later, using a \emph{view}. However, they can state \emph{axioms}
that any instance of the symbols in the parameter theory must satisfy.

For (a) above, the user provides a term of sort {\tt Machines}:

\begin{center}
\begin{small}
\begin{boxedminipage}{0.85\textwidth}
\begin{verbatim}
including MAP{NzNat,IOSize} * (sort Map{NzNat,IOSize} to Machines) .
\end{verbatim}
\end{boxedminipage}
\end{small}
\end{center}

\noindent
presumed to give a mapping from a prefix of the non-zero natural numbers,
isomorphic to the index set $J$, to pairs of numbers $(n_j,m_j)$, the number
of inputs and outputs of $M_j$, respectively. The sort for this pair of numbers
is {\tt IOSize}, defined with the following constructor

\begin{center}
\begin{small}
\begin{boxedminipage}{0.50\textwidth}
\begin{verbatim}
op _#_ : NzNat NzNat -> IOSize [ctor] .
\end{verbatim}
\end{boxedminipage}
\end{small}
\end{center}

With a term of sort {\tt Machines}, it is possible to generate the set of
parameters for the $\{M_j\}_{j\in J}$.  We simply iterate through the mappings,
creating a new \emph{parameter} for each $(n_j,m_j)$ pair requiring a view for
a theory 

\begin{center}
\begin{small}
\begin{boxedminipage}{0.75\textwidth}
\begin{verbatim}
  op smParams : Machines NzNat -> ParameterDeclList .
  eq smParams(MS, J) = 
      if $hasMapping(MS, J)
        then index('M, J) :: mkP(MS[J]) , smParams(MS, J + 1)
        else nil fi .

  op mkP : IOSize -> Qid .
  eq mkP(N # M) = -index(-index('SM, N), M) .
\end{verbatim}
\end{boxedminipage}
\end{small}
\end{center}

\noindent
The functions {\tt index} and {\tt -index} take a {\tt Qid} and a {\tt Nat} and
produce a new {\tt Qid}; for example, {\tt index('M, 1)} $=$ {\tt 'M1} and {\tt
-index('M, 1)} $=$ {\tt 'M-1}. The sort {\tt ParameterDeclList} is pre-defined
in the Maude prelude, in module {\tt META-MODULE}, using the following
constructors which are syntactically similar to the source-level representation
of parameters in a parameterized module

\begin{center}
\begin{small}
\begin{boxedminipage}{0.75\textwidth}
\begin{verbatim}
  op _::_ : Sort ModuleExpression -> ParameterDecl .
  op _,_ : ParameterDeclList ParameterDeclList 
      -> ParameterDeclList [ctor assoc id: nil prec 121] .
\end{verbatim}
\end{boxedminipage}
\end{small}
\end{center}

\noindent
The operator {\tt \$hasMapping} is pre-defined in the {\tt MAP} module; it
determines whether a given term has a mapping, and we use it to determine when
we are finished iterating through the $\vert J \vert$ modules.

We assume a set of theories, $\{\texttt{SM-$n_j$-$m_j$}\}_{j \in J}$, giving a
general interface for a synchronous machine with $n_j$ inputs and $m_j$
outputs; the general form of {\tt SM-$n_j$-$m_j$} is given in Figure
\ref{fig:sm}, it specifies the component input and output sorts, product types
for the input and output, projection functions, and a split transition
function. Figure \ref{fig:sminstance} shows how to instantiate a view of {\tt
SM-2-1} corresponding to a ``synchronous machine'' for the two xor gates given
above in Figure \ref{fig:circuit}. The {\tt TUPLE} module operation provides a
very general way to create product types (see \cite[\S
18.3.1]{clavel_2007_01}); it only works in Full Maude \cite[Part
II]{clavel_2007_01}, but by a slight abuse of notation we employ it in Figure
\ref{fig:sminstance}, and throughout this document, as if it can be used in
Core Maude \cite[Part I]{clavel_2007_01}; also, we call the projection
functions {\tt pi1}, \etc\, instead of {\tt p1\_} which is used in \cite[\S
18.3.1]{clavel_2007_01}. In addition, it is worth noting that while it would be
nice to use the {\tt TUPLE[\_]} module operation in the {\tt SM-$n$-$m$}
theories, parameterized theories are not currently allowed in Maude.

\begin{figure}[ht]
\begin{center}
\begin{boxedminipage}{0.55\textwidth}
\begin{small}
\begin{alltt}
fth SM-\(n\)-\(m\) is
  sorts Di Di-1 \(\dotsb\) Di-\(n\) .
  op `(_,\(\dotsb\),_`) : Di-1 \(\dotsb\) Di-\(n\) -> Di [ctor] .
  op pi1 : Di -> Di-1 .
  eq pi1(( X1:Di-1, \(\dotsb\) )) = X1:Di-1 .
  \(\dotso\)
  op pi\(n\) : Di -> Di-\(n\) .
  eq pi\(n\)(( \(\dotsb\), X\(n\):Di-\(n\) )) = X\(n\):Di-\(n\) .
  sort S .
  sorts Do Do-1 \(\dotsb\) Do-\(m\) .
  op `(_,\(\dotsb\),_`) : Do-1 \(\dotsb\) Do-\(m\) -> Do [ctor] .
  op pi1 : Do -> Do-1 .
  eq pi1(( X1:Do-1, \(\dotsb\) )) = X1:Do-1 .
  \(\dotso\)
  op pi\(m\) : Do -> Do-\(m\) .
  eq pi\(m\)(( \(\dotsb\), X\(m\):Do-\(m\) )) = X\(m\):Do-\(m\) .
  op delta1 : Di S -> S .
  op delta2 : Di S -> Do .
endfth
\end{alltt}
\end{small}
\end{boxedminipage}
\end{center}
\caption{The ``functional theory'' for a synchronous machine with $n$ inputs and $m$ outputs.}
\label{fig:sm}
\end{figure}

Consider again the example of Figure \ref{fig:circuit}. The corresponding {\tt
Machines} is given by

\begin{center}
\begin{small}
\begin{boxedminipage}{0.50\textwidth}
\begin{verbatim}
  1 |-> 2 # 1, 2 |-> 2 # 1, 3 |-> 2 # 1
\end{verbatim}
\end{boxedminipage}
\end{small}
\end{center}

\noindent
and the value produced by {\tt smParams} is

\begin{center}
\begin{small}
\begin{boxedminipage}{0.60\textwidth}
\begin{verbatim}
  'M1 :: 'SM-2-1 , 'M2 :: 'SM-2-1 , 'M3 :: 'SM-2-1 
\end{verbatim}
\end{boxedminipage}
\end{small}
\end{center}

\begin{figure}[ht]
\subfloat[Synchronous machine for an xor gate.]{
\input{inputs/sminstance1.tex}
}
\subfloat[View of the xor gate in the appropriate theory.]{
\input{inputs/sminstance2.tex}
}
\caption{Instantiating a synchronous machine for the xor gate using a view.}
\label{fig:sminstance}
\end{figure}

The values $n_e,m_e$, provided by the user and corresponding to (b) above,
determine the interface of the \emph{environment}, just as the $n_j,m_j$
determined the interface to the component synchronous machines.  Similar to the
{\tt SM-$n$-$m$} theories, we assume theories {\tt E-$n$-$m$} for the
environments. These are exactly the same as the {\tt SM-$n$-$m$} theories,
except that they omit the sort {\tt S} and the transition functions {\tt
delta1} and {\tt delta2}. Therefore, to generate the header for the module
giving the synchronous composition of a set of machines we can simply use

\begin{center}
\begin{small}
\begin{boxedminipage}{0.80\textwidth}
\begin{verbatim}
op scHeader : Machines IOSize -> Header .
eq scHeader(MS, N # M) = 
  'SC { smParams(MS, 1), E :: -index(-index('E, N), M) } .
\end{verbatim}
\end{boxedminipage}
\end{small}
\end{center}

The third component, (c) above, that we need to give is the \emph{wiring
diagram}. A wiring diagram is treated as a mapping between \emph{ports}, where
a port is defined as follows:

\begin{center}
\begin{small}
\begin{boxedminipage}{0.50\textwidth}
\begin{verbatim}
sort MIdx . subsort NzNat < MIdx . 
op e : -> MIdx [ctor] . 
sort Port . 
op `(_,_`) : MIdx NzNat -> Port [ctor] .
\end{verbatim}
\end{boxedminipage}
\end{small}
\end{center}

\noindent
Then, a wiring diagram is just (with a view for ports instantiated in the obvious way)

\begin{center}
\begin{small}
\begin{boxedminipage}{0.75\textwidth}
\begin{verbatim}
  including MAP{Port,Port} * (sort Map{Port,Port} to Wiring) .
\end{verbatim}
\end{boxedminipage}
\end{small}
\end{center}

To generate a module for the synchronous composition of machines $\{M_j\}_{j
\in J}$, and environment $E$, and a wiring diagram $\src$, we have a function

\begin{center}
\begin{small}
\begin{boxedminipage}{0.60\textwidth}
\begin{verbatim}
  op gensc : Machines IOSize Wiring -> Module .
\end{verbatim}
\end{boxedminipage}
\end{small}
\end{center}

\noindent 
where the arguments correspond to the three pieces of information, (a) -- (c)
respectively, above. The composed machine will be denoted by $\E$, following
the notation of \cite{meseguer_2010_01,meseguer_2009_01}.

The meta-level sort {\tt Module} in {\tt META-MODULE} contains the following
constructor for functional modules

\begin{center}
\begin{small}
\begin{boxedminipage}{0.90\textwidth}
\begin{alltt}
op fmod_is_sorts_.____endfm : Header ImportList SortSet SubsortDeclSet
  OpDeclSet MembAxSet EquationSet -> Module [ctor \(\dotsb\)].
\end{alltt}
\end{boxedminipage}
\end{small}
\end{center}

\noindent
The function {\tt gensc} is defined at the top level by instantiating each of
the arguments of the above operator as explained below

\begin{center}
\begin{small}
\begin{boxedminipage}{0.40\textwidth}
\begin{verbatim}
eq gensc(MS, E, W) =
    fmod
      scHeader  (MS, E, W) is 
      nil 
      sorts 
      scSorts   (MS, E, W) . 
      scSubsorts(MS, E, W) 
      scOpDecls (MS, E, W) 
      none 
      scEqs     (MS, E, W) 
    endfm .
\end{verbatim}
\end{boxedminipage}
\end{small}
\end{center}

The implementation of {\tt scHeader} is given above (albeit with the third
argument omitted, since it is unused and {\tt Wiring} had not been introduced).
For {\tt scSorts} we simply need to give \emph{names} for the relevant sorts of
$\E$: (1) the input type for the composed machine, $D^{\E}_i$, (2) the state
type, $S^{\E}$, and (3) the output type, $D^{\E}_o$.

\begin{center}
\begin{small}
\begin{boxedminipage}{0.60\textwidth}
\begin{verbatim}
op scSorts : Machines IOSize Wiring -> SortSet .
eq scSorts(MS, E, W) = 'Di^E ; 'Do^E ; 'S^E .
\end{verbatim}
\end{boxedminipage}
\end{small}
\end{center}

Let us jump ahead briefly to define the internal state of $\E$, \ie\ the
constructor for sort {\tt S\^{}E}, since it requires the notion of
\emph{internal node}, which we will need when defining {\tt scSubsorts}.
Let 
\begin{equation*}
N^{\E} = \{(j,m) \in J \times \N \mid \textit{$\exists \ (j',n) \in J \times \N$ s.t. $\src(j',n) = (j,m)$}\};
\end{equation*}
$N^{\E}$ is called the set of \emph{internal nodes}. Then, the state of $\E$ is
defined as
\begin{equation*}
\prod_{j \in J} S_j \times \prod_{(j,m) \in N} D^{j}_{o_{m}}
\end{equation*}
where $D^{j}_{o_{m}}$ denotes the sort of the $m^{th}$ output of machine $M_j$.
For example, the set of internal nodes for the circuit in Figure
\ref{fig:circuit} is $\{(1,1), (2,1)\}$, the outputs of the xor gates;
therefore we generate an {\tt OpDecl} for the state constructor as follows
(where the parameters are assumed to be the same as above)

\begin{center}
\begin{small}
\begin{boxedminipage}{0.95\textwidth}
\begin{verbatim}
(op `(_`,_`,_`,_`,_`) : 'M1$S 'M2$S 'M3$S 'M1$Do-1 'M2$Do-1 -> 'S^E [ctor] .)
\end{verbatim}
\end{boxedminipage}
\end{small}
\end{center}

\noindent
The general case is somewhat more tedious, but straightforward in the way
described above; for details, see our implementation \cite{katelman_2010_01}.

The component for subsorts, {\tt scSubsorts} is relatively simple to define,
but fraught with a subtle difficulty. To start with, we generate subsort
declarations for {\tt Di\^{}E} and {\tt Do\^{}E}, the inputs and outputs of the
composed module $\E$, respectively;

\begin{center}
\begin{small}
\begin{boxedminipage}{0.35\textwidth}
\begin{verbatim}
(subsort 'E$Do < 'Di^E .)
(subsort 'E$Di < 'Do^E .)
\end{verbatim}
\end{boxedminipage}
\end{small}
\end{center}

\noindent
We also need to give subsort declarations for input-output port matchings, for
example, to assert that the output sort of the xor gate $M_1$ is a subsort of
the first input of the and gate $M_3$

\begin{center}
\begin{small}
\begin{boxedminipage}{0.40\textwidth}
\begin{verbatim}
(subsort 'M1$Do-1 < 'M3$Di-1 .) 
\end{verbatim}
\end{boxedminipage}
\end{small}
\end{center}

\noindent
The subtle difficulty is that in Maude the semantics of {\tt subsort} is that
the sort on the left-hand side of the {\tt <} symbol is a \emph{proper} subsort
of the sort on the right-hand side. Therefore, when using this module for the
circuit in Figure \ref{fig:circuit}, one must guarantee that the input sorts to
the and gate are strict supersorts of the inputs. To maintain the genericity of
the parameterized module, this seems unavoidable, but one could go to a
\emph{less} generic solution where this problem would go away. In practice, we
have just added unit ($\star$) to the input type, \eg\

\begin{center}
\begin{small}
\begin{boxedminipage}{0.50\textwidth}
\begin{verbatim}
sort Bool+ . op * : -> Bool+ [ctor] .
\end{verbatim}
\end{boxedminipage}
\end{small}
\end{center}

Finally, we must generate operators and equations for the transition function
of $\E$. The operator declarations are straightforward

\begin{center}
\begin{small}
\begin{boxedminipage}{0.60\textwidth}
\begin{verbatim}
(op 'delta1^E : 'Di^E 'S^E -> 'S^E  [none].)
(op 'delta2^E : 'Di^E 'S^E -> 'Do^E [none].) .
\end{verbatim}
\end{boxedminipage}
\end{small}
\end{center}

\noindent
The equational definition of the above functions follows the description given
in \cite{meseguer_2010_01,meseguer_2009_01}. The details of the general case are too tedious to
present here, but in Figure \ref{fig:module} is shown the result of applying
{\tt gensc} to the appropriate arguments for the circuit in Figure
\ref{fig:circuit}. For full details, see our implementation
\cite{katelman_2010_01}.

\begin{figure}[ht]
\begin{center}
\begin{boxedminipage}{0.85\textwidth}
\begin{small}
\begin{verbatim}
fmod SC{M1 :: SM-2-1,M2 :: SM-2-1,M3 :: SM-2-1,E :: E-1-4} is
  sorts Di^E Do^E S^E .
  subsort E$Di < Do^E .
  subsort E$Do < Di^E .
  subsort E$Do-1 < M1$Di-1 .
  subsort E$Do-2 < M1$Di-2 .
  subsort E$Do-3 < M2$Di-1 .
  subsort E$Do-4 < M2$Di-2 .
  subsort M1$Do-1 < M3$Di-1 .
  subsort M2$Do-1 < M3$Di-2 .
  subsort M3$Do-1 < E$Di-1 .
  subsort M3$Do < E$Di .
  op `(_`,_`,_`,_`,_`) : M1$S M2$S M3$S M1$Do-1 M2$Do-1 -> S^E [ctor] .
  op pi1 : S^E -> M1$S  .
  op pi2 : S^E -> M2$S  .
  op pi3 : S^E -> M3$S  .
  op pi-1-1 : S^E -> M1$Do-1  .
  op pi-2-1 : S^E -> M2$Do-1  .
  op delta1^E : Di^E S^E -> S^E  .
  op delta2^E : Di^E S^E -> Do^E  .
  eq pi1   ((X1:M1$S,X2:M2$S,X3:M3$S,X4:M1$Do-1,X5:M2$Do-1)) 
        = X1:M1$S  .
  eq pi2   ((X1:M1$S,X2:M2$S,X3:M3$S,X4:M1$Do-1,X5:M2$Do-1)) 
        = X2:M2$S  .
  eq pi3   ((X1:M1$S,X2:M2$S,X3:M3$S,X4:M1$Do-1,X5:M2$Do-1)) 
        = X3:M3$S  .
  eq pi-1-1((X1:M1$S,X2:M2$S,X3:M3$S,X4:M1$Do-1,X5:M2$Do-1))
        = X4:M1$Do-1 .
  eq pi-2-1((X1:M1$S,X2:M2$S,X3:M3$S,X4:M1$Do-1,X5:M2$Do-1)) 
        = X5:M2$Do-1  .
  eq delta1^E(X:Di^E,Y:S^E) =
    ( delta1((pi1(X:Di^E),pi2(X:Di^E)),pi1(Y:S^E))
    , delta1((pi3(X:Di^E),pi4(X:Di^E)),pi2(Y:S^E))
    , delta1((pi-1-1(Y:S^E),pi-2-1(Y:S^E)),pi3(Y:S^E))
    , pi1(delta2((pi1(X:Di^E),pi2(X:Di^E)),pi1(Y:S^E)))
    , pi1(delta2((pi3(X:Di^E),pi4(X:Di^E)),pi2(Y:S^E)))
    ) .
  eq delta2^E(X:Di^E,Y:S^E) =
    ( pi1(delta2((pi-1-1(Y:S^E),pi-2-1(Y:S^E)),pi3(Y:S^E))) ) .
endfm
\end{verbatim}
\end{small}
\end{boxedminipage}
\end{center}
\caption{The parameterized module for the small circuit.}
\label{fig:module}
\end{figure}

\afterpage{\clearpage}

The last step is to instantiate the generated module with the appropriate
views.  For example, for the circuit in Figure \ref{fig:circuit} we use views
like the one given in Figure \ref{fig:sminstance}.

\begin{center}
\begin{small}
\begin{boxedminipage}{0.50\textwidth}
\begin{verbatim}
fmod USE-SC is
  including SC{Xor2,Xor2,And2,Env} .
endfm
\end{verbatim}
\end{boxedminipage}
\end{small}
\end{center}

One small difficulty in using the generated module is that it requires
capturing it, saving to a file, and then instantiating it with appropriate
views for the component machines and the environment. Of course, some small
changes need to be made for this to work, such as removing the quotes from the
quoted identifiers, but it is easy to write a small script for this purpose. It
would of course be more elegant to do this entirely inside of Maude, but
unfortunately the operations provided by {\tt META-LEVEL} make this difficult.
The essential problem is that because we are generating a \emph{parameterized}
module at the meta-level, to use it we need to \emph{instantiate} it, and that
requires generating a second meta-level module.  However, meta-level functions
such as

\begin{center}
\begin{small}
\begin{boxedminipage}{0.60\textwidth}
\begin{verbatim}
  op metaReduce : Module Term ~> ResultPair
\end{verbatim}
\end{boxedminipage}
\end{small}
\end{center}

\noindent
take as an argument just a \emph{single} module, and not a module \emph{set};
which we would need to capture both the generated parameterized module and its
instantiation. It may simply be better to generate a specialized, but
non-parameterized module instead of the parameterized one.

\section{Verifying LMST in the Presence of Failures using PALS}
\label{sec:lmst}

We now utilize the infrastructure described above to verify the correct
operation of the \emph{local minimum spanning tree protocol} \cite{li_2003_01}
in the presence of node failures. We begin with a brief introduction to the
protocol, what it aims to achieve and its basic operation, in Section
\ref{sub:lmst}. Then, we show how to verify the correctness of the LMST
protocol with respect to node failure. This entails showing how each individual
node is implemented as a synchronous machine of the kind described above
(Section \ref{sub:lmstToSMs}), composing the nodes, modeling the environment,
and performing the final verification (Section \ref{sub:experiment}).

\subsection{The LMST Protocol} 
\label{sub:lmst}

The purpose of a \emph{topology control protocol} is to define which nodes in an
ad-hoc wireless sensor network communicate with each other, and with what
transmission power they communicate. The goal is to minimize power consumption,
prolong network lifetime, and maximize data bandwidth while maintaining network
connectivity. 

In the case of the LMST protocol, a distributed, real-time algorithm is
employed whereby each sensor node periodically updates its own \emph{local
topology}. The local topology of a node is the set of neighboring nodes to
which it routes data. Each wireless node is a machine with internal quartz clock
timers, a memory for buffering messages, and a wireless transmitter which is
adjustable to different power levels. 

The periodic, real-time nature of the protocol is governed by a global constant
called the \emph{round time}, denoted $\rd$, and each node constantly employs
one of its timers, called the \emph{round timer}, to count the time between
round boundaries. When the round timer indicates that a new round has started
the node adjusts its local topology by changing its wireless transmission
strength. 

There are therefore two notions of a round, one \emph{global} and one
\emph{local}. A global round is any real-world interval $[t, t + \rd]$ where
$t$ is a multiple of $\rd$. A local round is an individual node's
\emph{perception} of the global, and is defined as any interval between
successive expirations of the node's local round timer, which may not keep
perfect time with respect to the real-world.

The protocol is then defined by what happens when the local round timer of a
node expires:

\begin{boxedminipage}{0.90\textwidth}
\begin{center}
\begin{minipage}{0.80\textwidth}
\begin{enumerate}
   \item
\label{step:1}
The node first broadcasts a message, called a \emph{hello message}, at
\emph{maximum transmission strength}. The hello message contains a unique
identifier of the node and its current physical location.  Hello messages are
buffered by any \emph{visible neighbor}, that is, any node within
wireless transmission range.

   \item
\label{step:2}
The node reads from its message buffer all hello messages received during the
previous round and distills from these a graph of its visible neighbors
weighted by distance.  

   \item
\label{step:3}
Taking the local graph of visible neighbors just distilled by the node, it then
calculates the minimum spanning tree of that graph.

   \item
\label{step:4}
The nodes in the local minimum spanning tree which are directly connected
(one-hop away) are selected to be the node's new \emph{neighbors}, meaning
those to which it will transmit data during this local round. The node then
scales its transmission power so that it can just reach the furthest of these
neighbors.

   \item
\label{step:5}
The node resets its round timer for $\rd$, and waits for the timer to expire.
\end{enumerate}
\end{minipage}
\end{center}
\end{boxedminipage}

As shown in \cite{li_2003_01}, LMST has a number of advantageous properties,
including low power usage, and a provably small number of neighbors for each
node, which reduces medium contention and increases bandwidth.  Furthermore, it
is also shown that LMST satisfies the crucial property of \emph{maintaining
network connectivity}.  That is, if the graph whose edges link the sensor nodes
within wireless reach of each other is connected, then the considerably smaller
subgraph computed by LMST is also connected.  

However, as described the protocol is somewhat idealized; it does not take into
account issues that must be faced in a real implementation such as medium
contention and node mobility. For the purpose of this paper, we ignore such
issues. For more information of formally analyzing the LMST protocol in a more
realistic setting, see \cite{katelman_2008_01}.

\subsection{LMST Nodes as PALS Synchronous Machines} 
\label{sub:lmstToSMs}


We now demonstrate an application of the PALS architecture to verify the
correctness of the LMST protocol in the presence of \emph{node failure}.  To do
this we use the infrastructure of Section \ref{sec:sc}, treating the wireless
nodes as individual synchronous machines and the environment as the determiner
of which nodes fail during a given step. Correctness is established by showing
that disconnectedness can only occur during a round when a node has failed.
This section treats just the construction of LMST nodes as \emph{synchronous
machines}, the modeling of the environment and formula we verify the system
against are described in Section \ref{sub:experiment}. Assume that all of the
definitions below go into a module {\tt LMST-NODE}, which we will use when we
instantiate the view associated with it at the end of this section.

For the sake of concreteness we consider a network with five nodes,
$N_1,\dotsc,N_5$, with an all-to-all topology. However, nodes will ignore any
hello message when it is outside the maximum range of the sending node.
Furthermore, nodes that fail will output a special token, {\tt nomsg},
indicating that no message was broadcast. 

\begin{center}
\begin{small}
\begin{boxedminipage}{0.50\textwidth}
\begin{verbatim}
pr    TUPLE[3]{NzNat,Nat,Nat} *
  ( sort Tuple{NzNat,Nat,Nat} to Msg 
  , op pi1 to id
  , op pi2 to xcoord
  , op pi3 to ycoord
  ) .

sort Msg+ . subsort Msg < Msg+ .
op nomsg : -> Msg+ [ctor] .
\end{verbatim}
\end{boxedminipage}
\end{small}
\end{center}

Given the all-to-all topology and the environment as described above, each node
will have five inputs: \emph{four} hello message lines, one each from each of
the other nodes, and \emph{one} from the environment determining if the node
fails during the current round. The input type for $N_1,\dotsc,N_5$ is
therefore given by

\begin{center}
\begin{small}
\begin{boxedminipage}{0.75\textwidth}
\begin{verbatim}
pr   TUPLE[5]{Msg+,Msg+,Msg+,Msg+,Status} * 
  (sort Tuple{Msg+,Msg+,Msg+,Msg+,Status} to RealInput) .

sort Di . subsort RealInput < Di .
op * : -> Di [ctor] .
\end{verbatim}
\end{boxedminipage}
\end{small}
\end{center}

\noindent
The additional constructor {\tt *} is necessary because of the semantics of
{\tt subsort} in Maude, as described above in Section \ref{sec:sc}. The sort
(with corresponding view) {\tt Status} contains two values associated with
constants, {\tt fail} and {\tt ok}. Note the sort name {\tt Di} corresponds to
a sort assumed in each of the {\tt SM-$n$-$m$}. This is convenient because it
allows us to avoid an explicit mapping when we eventually define the views for
each node $N_1,\dotsc,N_5$.

For the state of each node, we again have to consider the two cases where the
node is still running, or it has failed. If it is still running, it contains
all of the information it needs to send a hello message plus its current
routing table, which is a list of nodes to which it can route data through.

\begin{center}
\begin{small}
\begin{boxedminipage}{0.60\textwidth}
\begin{verbatim}
pr    TUPLE[4]{NzNat,Nat,Nat,NzNatList} *
  ( sort Tuple{NzNat,Nat,Nat,NzNatList} to NodeSt
  , op pi1 to id
  , op pi2 to xcoord
  , op pi3 to ycoord
  , op pi4 to routing
  ) .

sort S . subsort NodeSt < S .
op failed : -> S [ctor] .
\end{verbatim}
\end{boxedminipage}
\end{small}
\end{center}

Finally, the output type for each node just contains a single piece of
information for the hello message broadcast.

\begin{center}
\begin{small}
\begin{boxedminipage}{0.55\textwidth}
\begin{verbatim}
pr TUPLE[1]{Msg+} * (sort Tuple{Msg+} to Do) .
\end{verbatim}
\end{boxedminipage}
\end{small}
\end{center}

\noindent
Therefore, each of the nodes in the network will need to instantiate a view of
{\tt SM-5-1}, since each has five inputs and a single output.

We still need to define the transition function for each of the nodes
$N_1,\dotsc,N_5$. The transition function is exactly the same for each node

\begin{center}
\begin{small}
\begin{boxedminipage}{0.75\textwidth}
\begin{verbatim}
op delta1 : Di S -> S .
eq delta1(I, failed) = failed .
eq delta1(I, S     ) = 
    if pi5(I) == fail
      then failed
      else (id(S),xcoord(S),ycoord(S),routing'(I, S)) fi .

op delta2 : Di S -> Do .
eq delta2(I, failed) = (nomsg) .
eq delta2(I, S     ) = 
    if pi5(I) == fail
      then (nomsg)
      else ((id(S),xcoord(S),ycoord(S))) fi .
\end{verbatim}
\end{boxedminipage}
\end{small}
\end{center}

\noindent
where {\tt routing'} is defined according to the LMST algorithm given above in
Section \ref{sub:lmst} and \cite{li_2003_01} (for implementation details with
respect to our experiment, see \cite{katelman_2010_01}). Then, we can define a
node simply by giving instantiating a view with the above sorts and functions
as follows

\begin{center}
\begin{small}
\begin{boxedminipage}{0.60\textwidth}
\begin{verbatim}
view LMSTNode from SM-5-1 to LMST-NODE is
endv
\end{verbatim}
\end{boxedminipage}
\end{small}
\end{center}

\noindent
The body is empty because the module {\tt LMST-NODE} named its sorts and
operators using the same names and with the same signature as the {\tt SM-5-1}
theory.

\subsection{Verification of LMST using PALS} 
\label{sub:experiment}

With synchronous machines now for all of the nodes, we still must build the
composed machine, model the environment, and write an appropriate correctness
property. The first part is greatly eased by using the {\tt gensc} function
from Section \ref{sec:sc} with an all-to-all wiring diagram, and an appropriate
abstract environment, namely {\tt E-1-5}. Taking the environment from something
abstract to a concrete implementation is also straightforward, we essentially
just need a rule for each subset of nodes that can fail during a round. To do
this we first define an auxiliary function

\begin{center}
\begin{small}
\begin{boxedminipage}{0.60\textwidth}
\begin{verbatim}
op natToDi^E : Nat -> Di^E .
eq natToDi^E(X) = 
    ( if (X & (1 << 0)) > 0 then fail else ok fi
    , if (X & (1 << 1)) > 0 then fail else ok fi
    , if (X & (1 << 2)) > 0 then fail else ok fi
    , if (X & (1 << 3)) > 0 then fail else ok fi
    , if (X & (1 << 4)) > 0 then fail else ok fi
    ) .
\end{verbatim}
\end{boxedminipage}
\end{small}
\end{center}

\noindent
Where {\tt \&} is an operator for bit-wise and, and {\tt <<} is left-shift. As an example, to
generate an input where every node but the first one, $M_1$, fails, we simply
use {\tt natToDi\^{}E(1)} which evaluates to.

\begin{center}
\begin{small}
\begin{boxedminipage}{0.40\textwidth}
\begin{verbatim}
  (ok, fail, fail, fail, fail)
\end{verbatim}
\end{boxedminipage}
\end{small}
\end{center}

\noindent
The sort {\tt Di\^{}E} is just the automatically generated type via {\tt
gensc}, specifically (recalling from Section~\ref{sec:sc})

\begin{center}
\begin{small}
\begin{boxedminipage}{0.35\textwidth}
\begin{verbatim}
(subsort 'E$Do < 'Di^E .)
\end{verbatim}
\end{boxedminipage}
\end{small}
\end{center}

\noindent
which is just a $5$-tuple with every component of sort {\tt Status}, that is,
exactly the information we expect from the environment. Then, we add a
\emph{rule} to non-deterministically generate any possible output from the
environment (equivalently, \emph{input} to the device) at each step

\begin{center}
\begin{small}
\begin{boxedminipage}{0.60\textwidth}
\begin{alltt}
crl [fromEnv] : S => delta1^E(I, S) .
 if I,IS := possibleInputsSet
\end{alltt}
\end{boxedminipage}
\end{small}
\end{center}

\noindent
where {\tt possibleInputsSet} is a set all possible values of sort {\tt Di\^{}E}.
Using the function {\tt natToDi\^{}E} above, it is straightforward to generate

\begin{center}
\begin{small}
\begin{boxedminipage}{0.75\textwidth}
\begin{verbatim}
op genDi^EUpTo : Nat -> Set{Di^E} .
eq genDi^EUpTo(0   ) = natToDi^E(0   ) .
eq genDi^EUpTo(s(X)) = natToDi^E(s(X)) , genDi^EUpTo(X) .

op possibleInputsSet : -> Set{Di^E} .
eq possibleInputsSet = genDi^EUpTo(31) .
\end{verbatim}
\end{boxedminipage}
\end{small}
\end{center}

We still need to define a notion of correctness for LMST. At a high level we
say that the protocol is correct if the network always stays connected whenever
there are no new node failures during a round. The top-level LTL formula is
given by

\begin{center}
\begin{small}
\begin{boxedminipage}{0.70\textwidth}
\begin{verbatim}
op correct? : -> Formula .
eq correct? = O ([] no-new-failures? -> ((O connected?))) .
\end{verbatim}
\end{boxedminipage}
\end{small}
\end{center}

\noindent
which says, more precisely, that after the first time step it is always the
case that whenever the set of failing nodes is \emph{stable}, then during the
next round the network is \emph{connected}. The formula characterizing when
there are no new node failures is defined as

\begin{center}
\begin{small}
\begin{boxedminipage}{0.50\textwidth}
\begin{alltt}
op no-new-failures? : -> Formula .
eq no-new-failures? = 
    (failed?(1) <-> O failed?(1)) \(\wedge\)
    (failed?(2) <-> O failed?(2)) \(\wedge\)
    (failed?(3) <-> O failed?(3)) \(\wedge\)
    (failed?(4) <-> O failed?(4)) \(\wedge\)
    (failed?(5) <-> O failed?(5)) .
\(\dotso\)
eq S |= failed?(1) = pi1(S) == failed .
\(\dotso\)
eq S |= failed?(5) = pi5(S) == failed .
\end{alltt}
\end{boxedminipage}
\end{small}
\end{center}

\noindent
that is, that the failed proposition for each node is consistent between the
current state and the next state for every node individually. The {\tt
connected?} formula is more complicated, but essentially it traverses the
routing tables of all non-failed nodes to determine if there is a multi-hop
route from each non-failed node to every other one (see \cite{katelman_2010_01}
for details).

Finally, we can model check a $5$ node system against {\tt correct?} using
Maude's LTL model checker, showing that for the particular topology, LMST is
correct in the presence of node failure. Therefore, any \emph{asynchronous}
implementation of the system had through the PALS transformation would satisfy
the same notion of correctness.

\begin{center}
\begin{small}
\begin{boxedminipage}{0.85\textwidth}
\begin{verbatim}
Maude> red modelCheck(init, correct?) .
reduce in CHECK : modelCheck(init, correct?) .
rewrites: 317674 in 218ms cpu (226ms real) (1456678 rewrites/second)
result Bool: true
\end{verbatim}
\end{boxedminipage}
\end{small}
\end{center}

\section{Conclusions}
\label{sec:conclusions}

We have addressed the need to automatically support the synchronous composition
of abstract machines within Maude so that the PALS architecture can be
exploited for model checking purposes.  This is accomplished via
a meta-level module transformation in Maude that can automatically generate the
single abstract machine which is the composition of an ensemble of abstract
machines connected by a wiring diagram.  The transformation makes it easy to
verify a complex asynchronous DES system by model checking a much simpler
synchronous version where the user is responsible only for the individual
machine specifications and the wiring diagram.  

We have then illustrated how this transformation can be
applied to greatly simplify the formal verification of a key connectedness
property in the LMST topology control protocol for sensor networks.  As
explained in the introduction, a sensor network protocol such as LMST is an
example of a much broader class of distributed object-based DESs whose objects
only communicate with each other at pre-established times, and which change
their state at those times only as a result of the messages they then receive.
It would be quite useful to identify other examples of systems within this
category; also, the module transformation that we have presented could be
\emph{specialized to object-based systems} of this kind, so that it is not
necessary to specify such objects explicitly as abstract machines.

Besides the extension of the present work just outlined, much work remains
ahead.  For example, what is called transformation (2) in the Introduction
(passing from a synchronous model to its asynchronous PALS equivalent) should
also be automated at the meta-level, not for verification purposes, but for
purposes such as asynchronous design, code generation and also for system
emulation in physical time, when Real-Time Maude specifications are transformed
into actual real-time implementations.

\nocite{*}
\bibliographystyle{eptcs} 
\bibliography{rtrts_2010_camera}

\end{document}